\documentstyle[12pt,amssymb]{article}

\def\fnote#1#2{\begingroup\def\thefootnote{#1}\footnote{#2}\addtocounter{footnote}{-1}\endgroup}

\def\inbar{\vrule height1.5ex width.4pt depth0pt}
\def\IB{\relax{\rm I\kern-.18em B}}
\def\IC{\relax\,\hbox{$\inbar\kern-.3em{\rm C}$}}
\def\ID{\relax{\rm I\kern-.18em D}}
\def\IE{\relax{\rm I\kern-.18em E}}
\def\IF{\relax{\rm I\kern-.18em F}}
\def\IG{\relax\,\hbox{$\inbar\kern-.3em{\rm G}$}}
\def\IH{\relax{\rm I\kern-.18em H}}
\def\II{\relax{\rm I\kern-.18em I}}
\def\IK{\relax{\rm I\kern-.18em K}}
\def\IL{\relax{\rm I\kern-.18em L}}
\def\IM{\relax{\rm I\kern-.18em M}}
\def\IN{\relax{\rm I\kern-.18em N}}
\def\IO{\relax\,\hbox{$\inbar\kern-.3em{\rm O}$}}
\def\IP{\relax{\rm I\kern-.18em P}}
\def\IQ{\relax\,\hbox{$\inbar\kern-.3em{\rm Q}$}}
\def\IR{\relax{\rm I\kern-.18em R}}
\def\IT{\relax{\rm I\kern-.18em T}}
\def\ZZ{\relax{\sf Z\kern-.4em Z}}

\def\e{\epsilon}      
  \def\om{\omega}   \def\si{\sigma}

 \def\cH{{\cal H}}  
   
\def\cO{{\cal O}} \def\cP{{\cal P}}

\def\pfrak{{\mathfrak p}}
 \def\Hfrak{{\mathfrak H}}

\def\bIF{\bar \IF}   \def\bZZ{{\bar \ZZ}}

\def\Khat{{\hat K}}

\def\fnote#1#2{\begingroup\def\thefootnote{#1}\footnote{#2}\addtocounter
{footnote}{-1}\endgroup}
\def\beq{\begin{equation}}
\def\eeq{\end{equation}}
\def\bea{\begin{eqnarray}}
\def\eea{\end{eqnarray}}

\def\lleq#1{\label{#1}\eeq}
\let\nn=\nonumber
\def\tabroom{\hbox to0pt{\phantom{\Huge A}\hss}}
\def\notin{\ \hbox{{$\in$}\kern-.51em\hbox{/}}}

\def\lra{\longrightarrow}

  \def\E1Fq{E_1/\IF_q}

\def\IFp{{\IF_p}}  \def\IFpr{{\IF_{p^r}}}

\def\XIFp{{X/\IF_p}}  \def\XIFpr{{X/\IF_{p^r}}}

\def\rmH{{\rm H}} 
\def\rmN{{\rm N}}

\def\rmdR{{\rm dR}}     \def\rmdeg{{\rm deg}}
\def\rmdet{{\rm det}}   \def\rmdim{{\rm dim}}
   \def\rminf{{\rm inf}}
\def\rmord{{\rm ord}}   \def\rmmod{{\rm mod}}
\def\rmsign{{\rm sign}} \def\rmtr{{\rm tr}}

\def\rmGal{{\rm Gal}} \def\rmGL{{\rm GL}} 
\def\rmHW{{\rm HW}}   \def\rmSL{{\rm SL}} 

\def\notdiv{{\relax{~|\kern-.35em /~}}}

\begin{document}
\parindent=0pt
\hfill {\bf NSF-ITP-02-159}

\vskip 0.7truein

\centerline{\large {\bf THE SHIMURA-TANIYAMA CONJECTURE }}

\vskip .1truein

\centerline{\large {\bf AND CONFORMAL FIELD THEORY}}

\vskip 0.4truein

\centerline{\sc Rolf Schimmrigk$^1$\fnote{\dagger}{email:
netahu@yahoo.com} and Sean Underwood$^2$\fnote{\diamond}{email:
seancu@sowega.net}}

\vskip .4truein

\centerline{\it $^1$Kennesaw State University} \vskip .05truein
\centerline{\it 1000 Chastain Rd, Kennesaw, GA 30144}

\vskip .2truein

\centerline{\it $^2$Georgia Southwestern State University}\vskip
.05truein \centerline{\it 800 Wheatley St, Americus, GA 31709}

\vskip .6truein

\baselineskip=19pt

\centerline{\bf ABSTRACT:} \vskip .2truein
 \noindent
 The Shimura-Taniyama conjecture states that the Mellin
 transform of the Hasse-Weil L-function of any elliptic curve
 defined over the rational numbers is a modular form. Recent work
 of Wiles, Taylor-Wiles and Breuil-Conrad-Diamond-Taylor has
 provided a proof of this longstanding conjecture. Elliptic curves
 provide the simplest framework for a class of Calabi-Yau manifolds
 which have been conjectured to be exactly solvable. It is shown
 that the Hasse-Weil modular form determined by the arithmetic
 structure of the Fermat type elliptic curve is related in a
 natural way to a modular form arising from the character of a
 conformal field theory derived from an affine Kac-Moody algebra.

\vfill

{\sc PACS Numbers and Keywords:} \hfill \break Math:  11G25
Varieties over finite fields; 11G40 L-functions; 14G10  Zeta
functions; 14G40 Arithmetic Varieties \hfill \break Phys:
11.25.-w
Fundamental strings; 11.25.Hf Conformal Field Theory; 11.25.Mj
Compactification

\renewcommand\thepage{}
\newpage

\baselineskip=20.7pt
\parskip=.21truein
\parindent=0pt
\pagenumbering{arabic}

\section{Introduction}

Recent evidence suggests that the string theoretic nature of
spacetime can be illuminated by exploring the arithmetic structure
of the defining varieties. A first indication of the usefulness of
this technique is the identification of the quantum dimensions of
the chiral primary fields of exactly solvable string models as
certain units of a number field determined by the Hasse-Weil
L-function of the corresponding Calabi-Yau manifold. This result
provides a geometric characterization of the 'fine structure' of
the spectrum that goes beyond the usual identification of marginal
operators of the conformal field theory with elements of
cohomology groups of the variety \cite{rs01}.

Counting states is an issue that can be addressed in any of the
different formulations that have been applied to the problem of
understanding the relation between Calabi-Yau varieties and
conformal field theories, such as Landau-Ginzburg theory
\cite{m88, vw88, lvw89}, and the $\sigma$-theoretic approach
\cite{w93}. A question that so far has resisted efforts is how the
modularity of the two-dimensional conformal field theory is
encoded in the geometry of spacetime. More precisely, one wants to
understand how the characters of the underlying conformal field
theory are determined by the variety itself, and vice versa. It is
the purpose of this article to establish this relation in the
simplest case of toroidal Calabi- Yau compactifications.

Moral support for this investigation comes from the recent proof
of the more than three centuries old conjecture of Fermat's last
theorem. The basic ingredient of this proof is a conjecture first
put forward by Taniyama \cite{gs89}, sharpened by Shimura
\cite{gs71b}, and made more concrete by Weil \cite{w67} (ref.
\cite{lang} contains some remarks concerning the interesting
history of this conjecture). There are several different ways to
formulate this conjecture, but they all lead to a statement to the
effect that the arithmetically defined Hasse-Weil L-function of an
elliptic curve over the rational numbers is the Mellin transform
of a modular form of weight two with respect to some congruence
subgroup of $\rmSL(2,\ZZ)$. The geometric background for this
result is provided by Shimura's construction, which shows that the
Jacobian of an elliptic curve can be recovered as a factor of an
abelian variety determined by a modular form of weight two and
level $N$ \cite{gs71a}.

The Shimura-Taniyama conjecture has provided a focal point of much
work in arithmetic algebraic geometry over the last few decades.
It moved to center stage with Frey's observation \cite{f86} that
rational solutions of Fermat type plane curves can be used to
construct certain special types of semi-stable elliptic curves.
Frey argued that the resulting elliptic curves would in fact be so
special that they would contradict the Shimura-Taniyama
conjecture. Establishing the Shimura-Taniyama conjecture would
therefore finally prove Fermat's last theorem. Ribet's proof
\cite{kr90} of Frey's conjecture provided the key motivation for
Wiles' attempt to prove the conjecture, and hence Fermat's theorem
\cite{w95,tw95}. More recently the work of Wiles and Taylor-Wiles
has been extended to the full Shimura-Taniyama conjecture,
avoiding the requirement of semi-stability. Without any
constraints on the type of the elliptic curve the following result
has been proven in a sequence of papers with an increasing number
of authors \cite{d96,cdt99,bcdt01}. Denote by $\mathbb{Q}$ the
field of rational numbers, by $\IF_p$ the finite field of order
$p$, and by $E/\IF_p$ the reduced curve $E$ over the field
$\IF_p$. An elliptic curve $E$ is defined over $\mathbb{Q}$ if all
its coefficients are rational numbers. Set $q=e^{2\pi i\tau}$.

{\bf Theorem.}\cite{bcdt01} {\it Every elliptic curve $E$ over
$\mathbb{Q}$ is modular in the sense that there exists a modular
form $f=\sum_{n=1}^{\infty} a_n(f) q^n$ of weight 2 and some level
$N$, determined by the conductor of the elliptic curve, such that
the numbers} \beq a_p(E) = p+1-\#(E/\IFp)\eeq {\it defined by the
cardinalities $\#(E/\IF_p)$ of the curve at rational primes
$p\notdiv N$ are related to the coefficients $a_p(f)$ of the
modular form $f$ as} \beq a_p(E) = a_p(f).\eeq

Elliptic curves provide the simplest examples in a class of
Calabi-Yau manifolds which has been conjectured by Gepner to be
exactly solvable in terms of certain two-dimensional $N=2$
superconformal field theories \cite{g87}. Conformal field theories
are determined by partition functions whose ingredients are
holomorphic characters and multiplicity invariants which link the
holomorphic and anti-holomorphic sectors. In the context of
elliptic Calabi-Yau manifolds the most elementary exactly solvable
example is the cubic plane curve embedded in the projective plane
\beq C_3 = \{(z_0:z_1:z_2) \in \IP_2~|~ z_0^3 + z_1^3 + z_2^3
=0\}. \lleq{planecubic} The underlying conformal field theory of
this curve is thought to be derived from the affine SU(2) Lie
algebra at level $k=1$ \beq \IP_2 \supset C_3 ~\cong~\left({\rm
SU}(2)_{k=1,A_1}\right)_{\rm GSO}^{\otimes 3}, \lleq{cubicexact}
where $A_1$ signifies the diagonal invariant for the SU(2)
partition function, and GSO indicates the projection which
guarantees integral U(1)-charges of the states. The main
ingredient of the partition function of this theory is the string
function $c(\tau)$ which determines the character $\kappa(\tau)$
of the parafermionic theory at level $k=1$ via
$\kappa(\tau)=\eta(\tau)c(\tau)$, where $\eta(\tau) =
q^{1/24}\prod_{n=1}^{\infty}(1-q^n)$ is the Dedekind function. The
parafermionic theory in turn determines the $N=2$ superconformal
minimal models. Alternatively one can obtain the supersymmetric
theory via the Goddard-Kent-Olive coset construction.

The goal of this paper is to show that the theta functions
determined by the characters of the conformal field theory
determine the Hasse-Weil L-function in a simple way. More
precisely it is shown that the following holds. Denote by
$S_2(\Gamma_0(N))$ the set of cusp forms of the congruence group
of elements of $\rmSL(2,\ZZ)$ that are upper triangular mod $N$
\hfill \break \beq \Gamma_0(N) = \left\{\left(\matrix{a&b\cr
c&d\cr}\right) \in \rmSL_2(\ZZ) ~{\Big |}~\left(\matrix{a&b\cr
c&d\cr}\right) \sim \left(\matrix{*&*\cr 0&*\cr}\right)
~(\rmmod~N)\right\}. \eeq

{\bf Theorem.} {\it The Mellin transform of the Hasse-Weil
L-function $L_{\rmHW}(C_3,s)$ of the cubic elliptic curve $C_3
\subset \IP_2$ is a modular form $f_{\rmHW}(C_3,q) \in
S_2(\Gamma_0(27))$ which factors into the product} \beq
f_{\rmHW}(C_3,q) = \Theta(q^3)\Theta(q^9).\eeq {\it Here
$\Theta(\tau)=\eta^3(\tau)c(\tau)$ is the Hecke modular form
associated to the quadratic extension ${\mathbb Q}(\sqrt{3})$ of
the rational field $\mathbb{Q}$, determined by the unique string
function $c(\tau)$ of the affine Kac-Moody SU(2)-algebra at
conformal level $k=1$.}

This result provides a string theoretic origin of the Hasse-Weil
modular form of the plane cubic Fermat torus in terms of an
exactly solvable conformal field theory character determined by an
affine SU(2) Kac-Moody algebra.

Even though at present few exactly solvable points are known
explicitly, the Shimura-Taniyama conjecture shows that modularity
among Calabi-Yau curves is a common phenomenon, not restricted to
a few discrete points in moduli space. The link established in
this paper between geometrically defined modular forms and
conformal field theoretic modular forms, in combination with the
proofs of the Shimura-Taniyama conjecture, leads to the conjecture
that the space of exactly solvable Calabi-Yau varieties at central
charge $c=3$ is dense in the space of Calabi-Yau curves (see
\cite{gv02} for a different approach to this question). What is
needed is a better understanding of the conformal field theory
side of this relation.

The outline of this note is as follows. In Section 2 we describe
Artin's zeta function and the resulting Hasse-Weil L-function.
Section 3 contains a brief summary of some of the pertinent
aspects of $N=2$ superconformal field theories. Section 4 explains
the arguments that recover the conformal field theory modular form
from the variety, while Section 5 contains some remarks concerning
the inverse problem of reconstructing Calabi-Yau varieties from
conformal field theory. Section 6 describes an alternative point
of view in terms of the representation theory of the absolute
Galois group on the torsion points of the elliptic curve.

\section{The Hasse-Weil L-Function}

For reasons explained in \cite{rs01} it makes sense from a
physical perspective to combine the arithmetic information
contained at arbitrary prime numbers into a single object, the
Hasse-Weil L-function. Such a global object was first introduced
by Hasse in the late thirties (he suggested its investigation as a
dissertation topic to his student P. Humbert \cite{hh54}), and
later reformulated by Weil \cite{w50, w52}. The starting point for
the Hasse-Weil L-function of an algebraic curve $X$ is the local
congruent zeta function at a prime number $p$, defined by Weil
\cite{w49} as the generating series  \beq Z(\XIFp, t) \equiv
exp\left(\sum_{r\in \IN} \# \left(\XIFpr\right)
\frac{t^r}{r}\right) \eeq in terms of a formal variable $t$. In a
somewhat different formulation this function was introduced by
Artin \cite{a24} and F.K. Schmidt \cite{fks26}. This arrangement
of the cardinalities $\#(X/\IFpr)$ is motivated by the idea to
translate additive properties of these numbers into a
multiplicative structure of the generating function. It was first
shown by F.K. Schmidt \cite{fks31}\cite{hh33}\cite{hh34} that
$Z(\XIFp,t)$ is a rational function which takes the form \beq
Z(\XIFp,t) = \frac{\cP^{(p)}(t)}{(1-t)(1-pt)},\eeq where
$\cP^{(p)}(t)$ is a function whose degree is independent of the
prime $p$ and is given by the genus $g(X)$ of the curve,
$\rmdeg(X)=2g(X)$. It was later recognized that the structure
indicated by this simple expression holds quite generally in the
sense that the zeta function splits into factors determined by the
de Rham cohomology groups $\rmH_{\rmdR}(X)$ of a variety $X$ (see
\cite{pr97} for a beautiful historical introduction to this
subject).

More important from a physical perspective is the global zeta
function, obtained by setting $t=p^{-s}$ and taking the product
over all rational primes at which the variety has good reduction.
Denote by $S$ the set of rational primes at which $X$ becomes
singular and denote by $P_S$ the set of primes that are not in
$S$. The global zeta function can be defined as \beq Z(X,s) =
\prod_{p \in P_S} \frac{\cP^{(p)}(p^{-s})}{(1-p^{-s})(1-p^{1-s})}
= \frac{\zeta(s)\zeta(s-1)}{\prod_{p\in P_S} L_p(X,s)},
\lleq{globalzeta} where the local L-function has been defined as
\beq L_p(X,s) = \frac{1}{\cP^{(p)}(p^{-s})}, \lleq{local} and
$\zeta(s) = \prod_p (1-p^{-s})^{-1}$ is the Riemann  zeta of the
rational field $\mathbb{Q}$. It becomes clear from this way of
writing the zeta function that one of the advantages of defining
the L-function via eq. (\ref{globalzeta}) derives from the fact
that this expression separates the arithmetic of the variety from
that of the rational number field $\mathbb{Q}$.

The factors $\cP^{(p)}(t)$ can be analyzed in a number of
different ways. The most direct way is to expand Artin's form of
the zeta function and compare the coefficients with the expansion
of F.K.Schmidt's rational form of it.  Writing the polynomials
$\cP^{(p)}(t)$ at the good primes as \beq \cP^{(p)}(t)=
\sum_{i=0}^{2g} \beta_i(p)t^i,\eeq the first few coefficients
$\beta_i(p)$ expressed in terms of the $N_{r,p}= \#(X/\IFpr)$ are
given by \bea
\beta_0(p) &=& 1 \nn \\ \beta_1(p) &=& N_{1,p} - (p+1) \nn \\
\beta_2(p) &=& \frac{1}{2}\left(N_{1,p}^2 +
N_{2,p}\right) - (p+1)N_{1,p} + p \nn \\
  &\vdots &  \nn \\  \beta_{2g}(p) &=& p^g. \eea
At genus $g=1$ the zeta function is completely
  determined by $\beta_1(p)$.

Depending on the issues at hand it might be necessary to complete
this definition with factors coming from the bad and infinite
primes. The general structure of these factors is described in
\cite{s65}. For elliptic curves the discussion simplifies
considerably. For any prime $p$ the polynomials $\cP^{(p)}(t)$ for
elliptic curves can be written as \beq \cP^{(p)}(t) =
1+\beta_1(p)t + \delta(p)pt^2 \eeq with \beq \delta(p) =
\left\{\begin{tabular}{c l} 0 &if $p$ is a bad prime \\
1  &if $p$ is a good prime \tabroom \\
\end{tabular}\right\}. \eeq At the bad primes the precise
structure of the
coefficients $\beta_1(p)$ depends on the type of the singularity
\beq \beta_1(p) =
\left\{\begin{tabular}{c l} $\pm 1$ &if the singularity at $p$
is a node  \\
0  &if the singularity at $p$ is a cusp \tabroom \\
\end{tabular}\right\}. \eeq
Here the sign in the first case depends on whether the node is
split or non-split. The Hasse-Weil L-function can then be defined
as \beq L_{\rm HW}(X,s) = \prod_{p\in S}
\frac{1}{1+\beta_1(p)p^{-s}} \prod_{p \in P_S}
\frac{1}{1+\beta_1(p)p^{-s}+p^{2-2s}}.\eeq

Computing the cardinalities $N_{1,p}$ for the cubic curve $C_3$
given in eq. (\ref{planecubic}) explicitly allows to determine the
lower terms of the $q-$series. For the lower primes the
computation leads to the results in Table 1.
\begin{center}
\begin{tabular}{l| r r r r r r r r r r r}
Prime $p$    &2 &3 &5  &7  &11  &13    &17  &19  &23  &29  &31
\tabroom \\
\hline
$N_{1,p}$    &3 &4 &6  &9  &12  &9     &18  &27  &24  &30  &36
\tabroom \\
\hline
$\beta_1(p)$ &0 &0 &0  &1  &0   &$-5$  &0   &7   &0   &0   &4
\tabroom \\
\hline
\end{tabular}
\end{center}

{\bf Table 1.}{\it ~~The coefficients
$\beta_1(p)=N_{1,p}(C_3)-(p+1)$ of the elliptic cubic curve $C_3$
in terms of the cardinalities $N_{1,p}$ for the lower rational
primes.}

This leads to a Hasse-Weil series of the cubic elliptic curve \beq
L_{\rmHW}(C_3,s) = 1 - \frac{2}{4^s} - \frac{1}{7^s} +
\frac{5}{13^s} + \frac{4}{16^s} -\frac{7}{19^s} + \cdots \eeq

A standard maneuver then obtains from the Hasse-Weil L-series of
any elliptic curve $X$ \beq L_{\rmHW}(X,s) = \prod_{p\in S}
\frac{1}{1+\beta_1(p)} \prod_{p\in P_S}\frac{1}{1+\beta_1(p)
p^{-s}+p^{1-2s}} =\sum_{n=1}^{\infty} a_n n^{-s} \eeq an
associated $q-$expansion via the Mellin transform. This map
produces for a given $q$-series $f=\sum_n a_n q^n$ a series
$L(s)=\sum_n a_n n^{-s}$ via the integral \beq L(s) =
\frac{(2\pi)^s}{\Gamma(s)} \int_0^{\infty}f(iy)y^{s-1}dy.\eeq It
effectively replaces $n^{-s}\leftrightarrow q^n$, where $q=e^{2\pi
i \tau}$, and $\tau \in \Hfrak$ parametrizes the upper half plane.
This leads to the Hasse-Weil form \beq f_{\rmHW}(X,q) =
\sum_{n=1}^{\infty} a_n q^n \eeq associated to the Hasse-Weil
L-series. Applied to the Fermat cubic curve this leads to
 \beq f_{\rmHW}(C_3,q)
= q - 2q^4 - q^7 + 5q^{13} + 4q^{16} - 7q^{19} + \cdots
\lleq{hwmodform} Expansions like this often turn out to be useful
because of theorems which show that such functions are determined
uniquely by a finite number of terms.

This result raises a number of questions. First, we need to know
whether $f_{\rmHW}(C_3,q)$ is a modular form. That it is becomes
clear from the proof of the Shimura-Taniyama conjecture because
the Fermat cubic can be mapped into a Weierstrass form defined
over $\IQ$. It then follows from the result (\ref{hwmodform}) that
it is a cusp form (since $a_0=0$) and that it is normalized (since
$a_1=1$). We also need to know what its level and weight are.
Finally, we need to know whether it is a Hecke eigenform.

The weight of a Hecke eigenform can be read off directly from the
multiplicative properties of such a form. This can be seen as
follows. Consider the set $M_n$ of 2$\times$2 matrices over $\ZZ$
with determinant $n$. For $M=\left(\matrix{a&b\cr c&d\cr}\right)
\in M_n$ and a function $f$ on $\Hfrak$ define \beq (Mf)_k(\tau) =
\frac{\rmdet(M)^{k-1}}{(c\tau +d)^k}f(\tau).\eeq Define the Hecke
operators as \beq T_k(n) = \sum_{M\in M_n/\Gamma(1)}
(Mf)_k(\tau).\eeq Then \bea T_k(mn) &=&
T_k(m)T_k(n),~~~~~~~m,n~{\rm coprime}, \nn \\ T_k(p^{n+1}) &=&
T_k(p^n)T_k(p) - p^{k-1}T_k(p^{n-1}),~~~~p~{\rm prime}.\eea A
special operator has to be considered at primes $p$ which divide
the level $N$ of the form. This is the so-called Atkin-Lehner
operator \cite{al70}, for which no universal notation seems to
exist, but which is often denoted by $U_k(p)$ \beq U_k(p^n) =
(U_k(p))^n.\eeq For eigenforms of these operators the operator
structure translates into identical relation between their
coefficients $a_n$ \bea a_{mn}&=& a_ma_n~~~(m,n)=1 \nn \\
a_{p^{n+1}} &=& a_{p^n}a_p - p^{k-1}a_{p^{n-1}} \nn \\  a_{p^n}
&=& (a_p)^n,~~~~{\rm for}~p|N. \eea The Hasse-Weil form
$f_{\rmHW}(C_3,q)$ of the elliptic cubic curve satisfies these
relations with $k=2$, hence defines a normalized cusp Hecke
eigenform of weight two.

\section{Gepner Models}

The simplest class of N=2 supersymmetric exactly solvable theories
is built in terms of the affine SU(2) theory as a coset model
\begin{equation}  \frac{{\rm SU(2)}_k \otimes {\rm U(1)}_2}{{\rm
U(1)}_{k+2,{\rm diag}}}.\end{equation} Coset theories $G/H$ lead
to central charges of the form $c_G - c_H$, hence the
supersymmetric affine theory at level $k$  still has central
charge $c_k=3k/(k+2)$. The spectrum of anomalous dimensions
$\Delta^{\ell}_{q,s}$ and U(1)$-$charges $Q^{\ell}$ of the primary
fields $\Phi^{\ell}_{q,s}$ at level $k$ is given by
\begin{eqnarray} \Delta^{\ell}_{q,s} &=&
\frac{\ell (\ell +2)-q^2}{4(k+2)} + \frac{s^2}{8} \nonumber \\
Q^{\ell}_{q,s} &=& \frac{q}{k+2} - \frac{s}{2}, \end{eqnarray}
where $\ell\in \{0,1,\dots,k\}$, $\ell+q+s \in 2\mathbb{Z}$, and
$|q-s|\leq \ell$. Associated to the primary fields are characters
defined as \bea \chi^{\ell}_{q,s,u}(\tau,z) &=& e^{2\pi i u}
\rmtr_{\cH^{\ell}_{q,s}} q^{\left(L_0 - \frac{c}{24} \right)}
e^{2\pi i J_0} \nn \\ &=& e^{2\pi i u} \sum_{Q^{\ell}_{q,s},
\Delta^{\ell}_{q,s}} {\rm mult}\left(\Delta^{\ell}_{q,s},
Q^{\ell}_{q,s}\right) e^{2\pi i\left(\Delta^{\ell}_{q,s} -
\frac{c}{24}\right)+ 2\pi i Q^{\ell}_{q,s}}, \eea
 where the trace is to be taken over a
projection $\cH^{\ell}_{q,s}$ to a definite fermion number (mod 2)
of a highest weight representation of the (right-moving) $N=2$
algebra with highest weight vector determined by the primary
field. It is of advantage to express these maps in terms of the
string functions and theta functions, leading to the form
\begin{equation} \chi^{\ell}_{q,s}(\tau, z, u) = \sum
c^{\ell}_{q+4j-s}(\tau) \Theta_{2q+(4j-s)(k+2),2k(k+2)}(\tau,
z,u),
\end{equation} because it follows from this representation that
the modular behavior of the $N=2$ characters decomposes into a
product of the affine SU(2) structure in the $\ell$ index, and
into $\Theta$-function behavior in the charge and sector index.
The string functions $c^{\ell}_m$ are given by \beq
c^{\ell}_m(\tau) = \frac{1}{\eta^3(\tau)}
\sum_{\stackrel{\stackrel{-|x|<y\leq |x|}{(x,y)~{\rm
or}~(\frac{1}{2}-x,\frac{1}{2}+y)}}{\in
\ZZ^2+\left(\frac{\ell+1}{2(k+2)},\frac{m}{2k}\right)}} \rmsign(x)
e^{2\pi i \tau((k+2)x^2-ky^2)}, \eeq while the classical theta
functions $\Theta_{m,k}(\tau)$ are defined as  \beq
\Theta_{n,m}(\tau,z,u) = e^{-2\pi i m u} \sum_{\ell \in \ZZ +
\frac{n}{2m}} e^{2\pi i m \ell^2 \tau + 2\pi i \ell z}.\eeq It
follows from the coset construction that the essential ingredient
in the conformal field theory is the SU(2) affine theory.

It was suggested by Gepner fifteen years ago that exactly solvable
string compactifications obtained by tensoring several copies of
$N=2$ minimal models should yield, after performing appropriate
projections, theories that correspond in some limit to geometric
compactification described by Fermat type Calabi-Yau varieties.
The evidence for this conjecture was based initially mostly on
spectral information for all models in the Gepner class of
solvable string compactifications \cite{ls90, fkss90} and the
agreement of certain types of intersection numbers which allow an
interpretation as Yukawa couplings \cite{gkmr87, dgkm87, ss88,
rs90}. Alternative attempts to illuminate this surprising relation
were based on Landau-Ginzburg theories \cite{m88, vw88, lvw89} and
$\sigma$-models \cite{w93}. In the case of the Fermat cubic curve
these results suggest that there is an underlying conformal field
 theory of this elliptic curve that is described by the GSO
 projection of
 a tensor product of three models at conformal level $k=1$, as
 indicated in the Introduction.

Given the fact that certain types of elliptic curves lead to
modular forms, the question can be raised whether these forms are
related, in some way, to the modular forms that arise from the
conformal field theory. It is not clear a priori which of the
field theoretic quantities should be the correct building blocks
of the Hasse-Weil function, if any. What is clear is that none of
the characters by themselves can be sufficient, perhaps via some
polynomial expression, because their coefficients count
multiplicities of the primary states. Both, the characters of the
 affine SU(2) theory \beq
\chi^{\ell}(\tau,z,u) =
\sum_{\stackrel{n=-k+1}{n=\ell~\rmmod~2}}^k c^{\ell}_n(\tau)
\Theta_{n,k}(\tau,z,u), \eeq as well as the characters of the
parafermionic theory \cite{zf85} \beq \kappa^{\ell}_m(\tau) =
\eta(\tau) c^{\ell}_m(\tau), \eeq which provides the intermediate
step to $N=2$ minimal models \cite{zf86}, could in principle play
a r\^{o}le. In particular the string functions $c^{\ell}_m(\tau)$
of the $N=2$ characters would appear to be natural candidates
because they capture the essential interacting nature of the field
theory. At conformal level $k=1$ there is only one string
function, which we denote by $c(\tau)$, and which can be computed
to lead to the expansion \beq c(\tau) = q^{-1/24}(1+q+2q^2 + 3q^3
+ 5q^4 + 7q^5 + \cO(q^6)). \eeq It turns out that more important
than the string function itself is the associated SU(2) theta
function \beq \Theta^{\ell}_m(\tau) = \eta^3(\tau)
c^{\ell}_m(\tau).\eeq At arbitrary level $k$ these functions are
Hecke indefinite modular forms associated to quadratic number
fields determined by the level of the affine theory (ref.
\cite{kp84} contains background material). At level $k=1$ there is
a unique theta function, which we denote by $\Theta(\tau)$,
associated to the real quadratic extension ${\mathbb Q}(\sqrt{3})$
of the rational field $\mathbb{Q}$. Its expansion follows from the
string function expansion, resulting in \beq \Theta(q) = q^{1/12}
(1 - 2q - q^2 + 2q^3 + q^4 + 2q^5 + \cO(q^6)). \eeq This is a
modular form of weight one, which will emerge below as the
building block of the Hasse-Weil modular form $f_{\rmHW}(C_3,q)$
of the cubic elliptic curve.

\section{From Geometry to Modularity}

The comparison of the modular forms encountered so far shows that
there is no obvious relation between the geometric form
$f_{\rmHW}(C_3,q)$ and the string function $c(q)$, or the
associated theta function $\Theta(q)$ at (conformal) level 1. This
is not surprising for a number of reasons. The first is that
characters of the conformal field theory by themselves are not
useful because their coefficients count multiplicities, hence are
always positive.

Secondly, we expect the prospective Hasse-Weil modular form of an
elliptic curve to be of weight two. This is neither the weight of
the string function nor the weight of the theta function. The
theta function $\Theta(\tau)$ is a form of weight one, hence this
problem could easily be fixed by considering a product of two such
forms.

A further difference between the geometric and the conformal
 field theory forms is that the former have integral exponents,
  while the latter have rational exponents. Multiplying
two theta functions together will in general not automatically fix
this problem. The additional ingredient which serves as a useful
guide is the concept of the conductor. For an elliptic curve this
is a quantity which is determined both by the rational primes for
which the reduced curve degenerates, i.e. the bad primes, as well
as the degeneration type. Weil's important contribution to the
Shimura-Taniyama conjecture was his recognition that this
geometric conductor should determine the level of the modular form
given by the L-function induced series.

In the case of the Fermat cubic curve the bad prime is given by
$p=3$. This means that the level of the prospective number
theoretic modular form induced by the conformal field theory has
to be divisible by 3. The conductor of the curve can be computed
by first transforming the Fermat cubic into a Weierstrass form and
then applying Tate's algorithm \cite{t75}. Since we are interested
also in fields of characteristic 2 and 3, the usual (small)
Weierstrass form $y^2=4x^3+Ax+B$ is not appropriate. Instead we
have to consider the generalized Weierstrass form given by  \beq
E:~~~y^2 + a_1xy +a_3y =x^3 + a_2x^2 + a_4x +a_6,
\lleq{genweierstrass} where the unusual index structure indicates
the weight of the coefficients under admissible transformations
which preserve this form \beq (x,y)
~\mapsto~(u^2x+r,u^3y+u^2sx+t),\eeq with $r,s,t\in K$ and $u\in
K^*$ if $E$ is defined over $K$. Curves of this type can acquire
certain types of singularities when reduced over finite prime
fields $\IF_p$. The quantity which detects such singularities is
the discriminant \beq \Delta = \frac{c_4^3-c_6^2}{1728} \eeq where
\bea c_4 &=& b_2^2 - 24b_4 \nn
\\ c_6 &=& -b_2^2+ 36 b_2 b_4 - 216b_6,\eea with \bea
b_2 &=& a_1^2+4a_2 \nn \\ b_4&=&a_1a_3 + 2a_4 \nn \\
b_6 &=& a_3^2+4a_6. \eea The curve $E$ acquires singularities at
those primes $p$ for which $p|\Delta$. The singularity types that
can appear have been classified by Kodaira and N\'eron \cite{n64},
and are indicated by Kodaira's symbols. ${\rm I}_0$ describes the
smooth case, ${\rm I}_n, (n>0)$ involve bad multiplicative
reduction, and ${\rm I}_n^*, {\rm II}, {\rm III}, {\rm IV}, {\rm
II}^*, {\rm III}^*, {\rm IV}^*$ denote bad additive reduction.

The conductor itself depends on the detailed structure of the bad
fiber and the discriminant. Conceptually, it is defined as an
integral ideal of the field $K$ over which the elliptic curve $E$
is defined. By a result of Ogg \cite{o67} this ideal is determined
by the number $s_p$ of irreducible components of the singular
fiber at $p$ as well as the order $\rmord_p \Delta_{E/K}$ of the
discriminant $\Delta_{E/K}$ at $p$. In the present case the curve
is defined over the field $K=\mathbb{Q}$, hence the ring of
integers is a principal domain. The conductor can therefore be
viewed as a number defined by \beq N_{E/\mathbb{Q}} = \prod_{{\rm
bad}~p} p^{f_p},\eeq where the exponent $f_p$ is given by \beq f_p
= \rmord_p \Delta_{E/\mathbb{Q}} + 1 -s_p.\eeq

The Fermat cubic can be transformed into a Weierstrass form by
first choosing inhomogeneous coordinates and setting \beq
\frac{x}{z} \mapsto -\frac{3u}{v},~~~~\frac{y}{z} \mapsto
\frac{9-v}{v}.\eeq This leads to the form $v^2-9v=u^3-27$. This
result can be transformed further by completing the square and
introducing the variables $x=u$ and $y=v-5$, leading to the affine
curve \beq y^2 + y = x^3 - 7, \eeq with discriminant $\Delta
=-3^9$ and $j-$invariant $j=0$. The singular fiber resulting from
Tate's algorithm is of Kodaira type IV$^*$ with $s_3=7$
components, leading to the conductor $N=27$.

\begin{center}
{\thicklines
 \begin{picture}(100,50)
        \put(20,10){\line(1,0){120}}
        \put(50,0){\line(0,1){50}}
        \put(40,20){\line(1,1){25}}
        \put(80,0){\line(0,1){50}}
        \put(65,35){\line(1,0){30}}
        \put(110,0){\line(0,1){50}}
        \put(120,20){\line(-1,1){25}}
 \end{picture}} \break
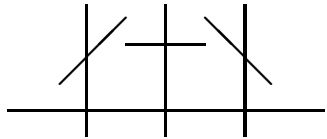
 {\bf Figure 1.}{\it ~~The Kodaira type ${\rm IV}^*$ type
singularity, indicating the intersection properties.}
\end{center}

Combining the weight consideration with the integrality condition,
as well as the conductor computation, suggests to look for modular
forms of the type $\Theta(3a\tau)\Theta(3b\tau)$, where $a,b$ are
integers such that $(a+b)=4$. This leads to the ansatz $f_1(\tau)
= \Theta(3\tau) \Theta(9\tau)$ as a candidate modular form at
conformal level $k=1$. Expanding this form gives
\bea f_1(\tau) &=& \Theta(3\tau)\Theta(9\tau) \nn \\
 &=& q - 2q^4 - q^7 + 5q^{13} + 4q^{16} - 7q^{19} - 5q^{25}
 + 2q^{28} - 4q^{31} + \cdots. \eea Comparing this conformal field
 theoretic form with the form (\ref{hwmodform}) shows
 complete agreement. This establishes a relation between the
 geometrically determined modular form derived from the Hasse-Weil
 L-function and a number theoretic modular form of the quadratic
 field ${\mathbb Q}(\sqrt{3})$ derived from an affine Kac-Moody
 algebra.

From the considerations of Section 2 we know that the form
$f_1(\tau)$ is a normalized cusp Hecke eigenform of weight two. In
order to complete the identification of this form it is useful to
recognize that the string function at conformal level one is given
by the inverse of the Dedekind eta-function. Hence the theta
function is given by the square of the $\eta$-function. This
allows us to determine the (modular) level of the form by
considering the level of the Fricke involution on the set of cusp
forms of level $N$ induced by the matrix \beq w_N =
\left(\matrix{0&-1\cr N&0\cr}\right).\eeq An eigenform $f$ of
weight 2 and level $N$ transforms as \beq
f\left(-\frac{1}{N\tau}\right) = \pm N\tau^2 f(\tau).\eeq For the
series $f_1(\tau)$ the Fricke involution leads to the result \beq
f_1\left(-\frac{1}{27\tau}\right) = 27 \tau^2 f_1(\tau).\eeq Hence
$f_1$ is a form at level 27, $f_1(\tau) =
\Theta(3\tau)\Theta(9\tau) \in S_2(\Gamma_0(27))$.

This leaves the question whether there are other forms of this
type. The dimension of $S_2(\Gamma_0(N))$ can be determined via
the theory of modular curves $X_0(N)$ \cite{gs71a}. These are
objects defined by quotients \beq X_0(N) =
\Hfrak^*/\Gamma_0(N),\eeq where $\Hfrak^* = \Hfrak \cup \mathbb{Q}
\cup \infty$ and $\Hfrak$ is the upper half plane. Each form $f\in
S_2(\Gamma_0(N))$ defines a differential form $\om_f = 2\pi i
f(\tau) d\tau$ on $X_0(N)$, hence the dimension of
$S_2(\Gamma_0(N))$ is given by the genus of the curve \beq \rmdim
S_2(\Gamma_0(N)) = g(X_0(N)).\eeq The latter is determined
completely by the index $\mu(N)$ of $\Gamma_0(N)$ in
$\Gamma(1)=\rmSL(2,\ZZ)$, its number of elliptic points of order 2
and 3, $\nu_2(N)$ and $\nu_3(N)$, and the number of $\Gamma_0(N)$
inequivalent cusps $\nu_{\infty}(N)$. One has the following result

{\bf Theorem.} (\cite{gs71a}) {\it The genus of $X_0(N)$ is given
by} \beq g(X_0(N)) = 1+ \frac{\mu(N)}{12} - \frac{\nu_2(N)}{2} -
\frac{\nu_3(N)}{2} - \frac{\nu_{\infty}}{2}.\eeq {\it Here the
index is given by} \beq \mu(N) = [\Gamma(1): \Gamma_0(N)] = N
\prod_{p|N} \left(1+\frac{1}{p}\right),\eeq {\it while the number
of elliptic points of order 2 and 3 are given by} \bea \nu_2(N)
&=& \left\{\begin{tabular}{l l} 0 &if
$N$ is divisible by 4 \tabroom \\
$\prod_{p|N}\left(1+\left(\frac{-1}{p}\right)\right)$  &otherwise
\tabroom \\
\end{tabular} \right\} \nn \\
\nu_3(N) &=& \left\{\begin{tabular}{l l} 0 &if
$N$ is divisible by 9 \tabroom \\
$\prod_{p|N}\left(1+\left(\frac{-3}{p}\right)\right)$  &otherwise
\tabroom \\
\end{tabular} \right\},
\eea {\it where the symbol $\left(\frac{}{p}\right)$ denotes the
quadratic residue symbol defined as} \bea
\left(\frac{-1}{p}\right) &=& \left\{\begin{tabular}{r l} 0 &if
$p=2$ \tabroom \\ 1  &if $p\equiv 1~\rmmod~4$ \tabroom \\ $-1$ &if
$p\equiv 3~\rmmod~4$ \tabroom \\ \end{tabular}\right\} \nn \\
\left(\frac{-3}{p}\right) &=& \left\{\begin{tabular}{r l} 0 &if
$p=3$ \tabroom \\ 1  &if $p\equiv 1~\rmmod~3$ \tabroom \\ $-1$ &if
$p\equiv 2~\rmmod~3$ \tabroom \\ \end{tabular}\right\} \eea {\it
The number of cusps is given by} \beq \nu_{\infty}^0(N) =
\sum_{0<d|N} \phi((d,N/d)).\eeq {\it Here $(d,N/d)$ denotes the
greatest common divisor and $\phi(n)$ is the Euler totient
function.}

 Computing the genus of the curve $X_0(27)$ then shows that the
space $S_2(\Gamma_0(27))$ is one-dimensional, and that the form
$f_1(\tau)$ is the unique generator (up to constants).

A priori these two series could be different at higher than
computed order but, as mentioned before, general results by
Faltings and Serre show that modular forms which agree to a
sufficiently high, but finite, order, actually coincide. For
modular forms of congruence groups $\Gamma_0(N)$ there is an
explicit result which determines the congruence in a simple way
\cite{s87}. For the situation of relevance here, this can be
formulated as follows.

{\bf Theorem.}(Sturm)~{\it Let $f=\sum_na_nq^n$ and $g=\sum_n
b_nq^n$ be modular forms of weight $k$ with coefficients in the
ring of integers $\cO_K$ of some number field $K$. For prime
ideals  $\pfrak \subset \cO_K$ and forms $f$ define } \beq
\rmord_{\pfrak}(f) = \rminf\{n~|~\pfrak \notdiv a_n\}.\eeq {\it
Then, if} \beq\rmord_{\pfrak}(f-g) >
\frac{k[\Gamma_0(1):\Gamma_0(N)]}{12} \eeq {\it one finds that
$\rmord_{\pfrak}(f-g)=\infty$, i.e. $\pfrak|(a_n-b_n)$ for all
$n$.}

This completes the proof of the theorem formulated in the
Introduction.

In the present case of elliptic curves constructive information is
available in the form of the Eichler-Shimura theory. The following
section describes how this framework fits into the general scheme
of a conformal field theoric analysis of geometric
compactifications via arithmetic methods.

\section{From Conformal Field Theory to Geometry}

Perhaps the most important problem in the context of string
compactification is to invert the analysis described in the
previous sections, and to ask how to construct a geometric target
space from the data provided by the conformal field theory. This
section describes the first steps in this direction. Our focus
will mostly be on elliptic curves. This does not mean that these
considerations are irrelevant for higher dimensional varieties.
Examples where elliptic curves span part of the cohomology are
threefolds such as transverse hypersurfaces of degree twelve
embedded in $\IP_{(1,2,3,3,3)}$. Here the singular set is given by
the torus $C_3 \subset \IP_2$ and its resolution contributes to
the spectrum of the threefold. This resolution itself is
determined in part by the cohomology of the curve $C_3$ and
therefore the arithmetic of the elliptic curve becomes part of the
arithmetic structure of the variety.

Conformal field theories can more usefully be thought of as
determining pieces of cohomology, or more precisely, the motivic
structure of the variety. Specifying the target motive is part of
the external data that needs to be supplied to the conformal field
theory. In the previous sections we have seen that starting from a
variety it is possible to construct functions on the upper half
plane that exhibit modular behavior, at least sometimes. In the
case of elliptic curves over the rationals this has now been shown
to always be the case, confirming Taniyama's intuition beyond his
own expectations.

It is an open question at present which modular forms, or more
generally, automorphic forms, are induced by motives. The
structure that has emerged from the analysis of the congruent zeta
function shows that a relation between modular forms and geometric
objects very likely can be established only by considering certain
subclasses of forms. (1) The form $f=\sum_n a_n q^n$ should be a
cusp form, i.e. $a_0=0$. (2) $f$ should be normalized, i.e.
$a_1=1$. (3) $f$ should be modular at some level $N$. (4) The form
$f$ should be a Hecke eigenform.

The expectation that these properties might suffice to guarantee a
geometric origin of a modular form is suggested by the analysis of
Atkin and Lehner.  Denote the set of forms of weight 2 and level
$N$ by $S_2(N)$, and define a newform as an element in this space
that is not determined by a level $N'$ that is a divisor of $N$.

{\bf Theorem.} (Atkin-Lehner) {\it Let $S_k(\Gamma_0(N)) \ni
f=\sum_{n=1}^{\infty} a_nq^n$ be a normalized cusp form which is a
newform and is an eigenvector for all Hecke operators $T_k(n)$.
Then} \beq L(f,s) = \prod_{\stackrel{p {\rm prime}}{p|N,
p^2\notdiv N}} \frac{1}{1+\lambda(p)p^{\frac{k}{2}-1-s}}
\prod_{\stackrel{p {\rm prime}}{p\notdiv N}} \frac{1}{1-a_pp^{-s}
+ p^{k-1-2s}},\eeq {\it where $\lambda(p)=\pm 1$.}

Comparing this result with the definition of the Hasse-Weil
L-series shows that it is this class of modular forms which admits
a possible geometric interpretation.

The strategy adopted here to go from such modular forms to
geometry is based on results by Eichler and Shimura
\cite{me54,gs58}. Given a cusp Hecke eigenform $f$ of weight two
and level $N$ the Eichler-Shimura construction leads to an abelian
variety $A_f$ whose dimension is determined by the degree of the
field extension $K={\mathbb Q}(\{a_n\})$ defined by the
coefficients of the modular form \cite{gs59} \beq
\rmdim_{\mathbb{C}} A_f = [K:\mathbb{Q}].\lleq{dimAf} The
importance of $A_f$ derives from the fact that its arithmetic
properties are determined by the modular form in the sense that
the Mellin transform of its Hasse-Weil L-function $L(A_f,s)$ is
determined by the L-series of the modular form $f$. In more
detail, the concrete construction of $A_f$ proceeds by recovering
it as a factor in the Jacobian of the modular curve $X_0(N)$
defined as the compactification of the affine curve \beq Y_0(N) =
\Hfrak/\Gamma_0(N), \eeq  where $\Hfrak$ is the upper half plane.

Intuitively, this abelian factor can be viewed as the subvariety,
or factor, of the Jacobian of the modular curve spanned by the set
$\{f^{\si}\}$ of conjugate forms obtained from the set $\{\si:
K\lra \mathbb{C}\}$ of embeddings of the coefficient field $K$.
Associated to a modular form $f$ of weight 2 and level $N$ on the
upper half plane ${\mathfrak H}$ is a holomorphic differential
$\om_f = 2\pi i f(z) dz$, which descends to the quotient Riemann
surface $\Gamma_0(N)\backslash{\mathfrak H}$ and extends to the
compactification $X_0(N)$ over ${\mathbb Q}$, defining a 1$-$form
$\om_f$ on the modular curve $X_0(N)$.

If $f$ has rational coefficients then (\ref{dimAf}) shows that the
abelian variety is an elliptic curve $E$. The results of Eichler
and Shimura can further be used to establish that the Hasse-Weil
$L-$function $L(E,s)$ associated to $f$ agrees with the $L-$series
of $f$ for almost all primes. The construction of Eichler-Shimura
provides a map \beq \Phi_f: ~\Gamma_0(N) \lra \mathbb{C} \eeq that
annihilates the elliptic and parabolic points. If $f$ is not only
a newform but also a Hecke eigenform such that the Hecke
eigenvalues are integers then $\Phi_f$ determines a lattice and
the resulting elliptic curve $E$, the modular curve $X_0(N)$, and
the map \beq X_0(N) \lra E \eeq are defined over $\mathbb{Q}$.
Igusa \cite{i73} improved this result by showing that the only
possible exception are the bad primes dividing the level $N$. It
was finally shown by Carayol that indeed the arithmetic agreement
holds for all primes and therefore one finally has the following
result.

{\bf Theorem.}\cite{c86}~{\it Let $f\in S_2(\Gamma_0(N))$ be a
normalized newform with integral coefficients in ${\mathbb Z}$ and
let $E$ be the elliptic curve associated to $f$ via the
Eichler-Shimura construction. Then}
 \beq L(E,s) = L(f,s) \eeq
{\it and $N$ is the conductor of $E$.}

It is not obvious that the curve $E$ obtained in this way should
have the same arithmetic as the Fermat curve which was the
starting point of the considerations above. It turns out, however,
that both the Fermat curve $C_3$ and the abelian variety $A_f$ are
on equal footing as far as their arithmetic properties are
concerned. The concept that captures what might be called the
'arithmetic equivalence' of elliptic curves, and more generally of
abelian varieties, is that of an isogeny. A non-constant analytic
map $E \lra E'$ between two elliptic curves $E$ and $E'$ is called
an isogeny if it takes the distinguished point $O$ of $E$, given
by the zero of the algebraic group structure, into the
corresponding point $O'$ of $E'$. If such a map exists then $E$
and $E'$ are said to be isogenous. Two isogenous elliptic curves
over $\mathbb{Q}$ have the same primes of bad reduction
\cite{st68}, and furthermore for each good prime their
cardinalities agree, $\#(E/\IF_p) = \#(E'/\IF_p)$.  An important
result by Faltings, proving a conjecture of Tate, shows that the
converse holds.

{\bf Theorem.} (Tate, Faltings \cite{gf83}) \hfill \break {\it Two
elliptic curves $E,E'$ over the rational field $\IQ$ which have
the same L-function are isogenous.}

These results indicate that in the context of trying to understand
conformal field theoretical aspects of string compactifications
via the arithmetic structure of the varieties we should not
consider individual varieties as objects corresponding to the
conformal field theory. Instead it is more useful to think in
terms of equivalence classes of varieties, where the equivalence
relation is given by the concept of isogeny.

\section{Representation theoretic framework}

There is an alternative framework which provides a tool to analyze
the arithmetic structure of algebraic varieties, and which allows
to think somewhat differently about the deeper issues involved in
the relation between Calabi-Yau varieties and affine Kac-Moody
algebras. This formulation involves representations of the
absolute Galois group $\rmGal({\bar {\mathbb Q}}/{\mathbb Q})$,
defined as the group of automorphisms of the algebraic closure
${\bar {\mathbb Q}}$ of the rational field that leaves ${\mathbb
Q}$ fixed, and therefore relates to a question raised in
\cite{m98}. Moore observed that the Galois group of the class
field $\Khat$ of a field $K$ acts on the vector multiplet
attractor moduli and asked whether the absolute Galois group
$\rmGal({\bar {\mathbb Q}}/{\mathbb Q})$ plays a role in some
compactifications. The following considerations show such an
application in a physical context.

The group $\rmGal({\bar {\mathbb Q}}/{\mathbb Q})$ can be viewed
as a limit of the Galois groups $\rmGal(K/{\mathbb Q})$ where $K$
runs through all possible finite extensions of ${\mathbb Q}$ in
${\bar {\mathbb Q}}$. Any element of $\rmGal({\bar {\mathbb
Q}}/{\mathbb Q})$ determines a system $\si$ of automorphisms
$\si_K$ in each of the individual Galois groups $\rmGal(K/{\mathbb
Q})$. The $\si_K$ are compatible in the sense that if one has two
extensions $K,L$ of ${\mathbb Q}$ such that $K\subset L$ then the
automorphisms $\si_K \in \rmGal(K/{\mathbb Q})$ and $\si_L \in
\rmGal(L/{\mathbb Q})$ are such that $\si_L$ restricts to $\si_K$,
i.e. $\si_K = \si_L|_K$.

The goal in number theory is to understand the absolute Galois
group $\rmGal({\bar {\mathbb Q}}/{\mathbb Q})$ via its
$n$-dimensional continuous representations \beq \rho: \rmGal({\bar
{\mathbb Q}}/{\mathbb Q}) ~\lra~\rmGL(n,K),\eeq where $K$ is a
(topological) field which can e.g. be the complex field
$\mathbb{C}$, the $p-$adic field ${\mathbb Q}_p$ or some extension
of it, or it can be some finite field. The latter case turns out
to be of interest in the present context.

For any elliptic curve $E$ representations of the absolute Galois
group can be constructed from the group $E[n]$ of torsion points ,
i.e. the subgroup of points of the group $E({\bar {\mathbb Q}})$
for which $nx=0$. The underlying group operation is written
additively, $P_1+P_2$, where the sum of two points is defined as
the intersection point of the line connecting the points $P_1,P_2$
with the elliptic curve (see e.g. \cite{k92}). If the curve is
described by the generalized Weierstrass form
(\ref{genweierstrass}), the line $y=mx+b$ connecting two points
$P_1=(x_1,y_1)$ and $P_2=(x_2,y_2)$ is determined by

\beq m =\left\{\begin{tabular}{l l}
 $\frac{y_2-y_1}{x_2-x_1}$,  &if $P_1\neq P_2$ \\
 $\frac{3x_1^2+2a_2x_1+a_4-a_1y_1}{2y_1+a_1x_1+a_3}$, &if
 $P_1=P_2$ \tabroom \\ \end{tabular} \right\} \eeq
 and
\beq b =\left\{\begin{tabular}{l l}
 $\frac{y_1x_2-y_2x_1}{x_2-x_1}$,  &if $P_1\neq P_2$ \\
 $\frac{-x_1^3+a_4x_1+2a_6-a_3y_1}{2y_1+a_1x_1+a_3}$, &if
 $P_1=P_2$ \tabroom \\ \end{tabular} \right\}. \eeq

The intersection point $P_3=(x_3,y_3)$ of the line with the curve
is then given by \bea x_3&=&m^2+a_1m-a_2-x_1-x_2 \nn \\ y_3&=&
 -(m+a_1)x_3 -b-a_3 \eea in both the chord case ($P_1\neq P_2$)
 and the tangent case ($P_1=P_2$).

The group $E[n]$ of points of order $n$ is a free
$\ZZ/n\ZZ$-module of rank 2,
 i.e. $E[n] \cong \ZZ/n\ZZ \times \ZZ/n\ZZ$. It is possible to
 define an action of the absolute Galois group on $E[n]$ via a
 homomorphism \beq \rho_n:~
 {\rm Gal}({\bar {\mathbb Q}}/{\mathbb Q}) ~\lra ~ {\rm Aut} (E[n])
 \cong {\rm GL}(2,\ZZ/n\ZZ).\eeq The image
 $\rho_n(\rmGal(\bar{\mathbb{Q}}/\mathbb{Q}))$ in
 $\rmGL(2,\ZZ/n\ZZ)$ can be thought of as the Galois group
 $\rmGal(K[n]/\mathbb{Q})$ of the field extension $K[n]$ obtained by
 adjoining the coordinates of the points of $E[n]$ to the
 rationals. More precisely, one has the following

{\bf Theorem.}\cite{jps68} {\it The $x$ and $y$ coordinates of the
points of $E[n]$ have algebraic values. If $K[n]$ is the field
extension obtained from ${\mathbb Q}$ by adjoining all coordinates
of these points then $K[n]$ is a Galois extension and the
representation $\rho_n$ factors through $\rmGal({\bar {\mathbb
Q}}/{\mathbb Q})$, i.e.} \beq \rmGal({\bar {\mathbb Q}}/{\mathbb
Q}) ~\lra ~\rmGal(K[n]/{\mathbb Q}) ~\lra ~{\rm GL}(2,\ZZ/n\ZZ),
\eeq {\it where the map on the left is the canonical surjection
and the map on the right is injective.}

 The interesting aspect of these representations is that they can
 be related to the arithmetic of the variety discussed in previous
 sections. Roughly, this can be outlined as follows. From the
 normal extension $K[n]/\mathbb{Q}$ one obtains a map \beq \rho:
\rmGal(K[n]/ {\mathbb Q})~\lra ~ {\rm GL}(2,\ZZ/n\ZZ).\eeq The
subfield of $K[n]$ whose Galois group is
$\rho^{-1}(\rmSL(2,\ZZ/n\ZZ)) \subset \rmGal(K[n]/\mathbb{Q})$ is
the cyclotomic field ${\mathbb Q}(\mu_n)$. The basic idea is to
associate to a rational prime $p$ a class of conjugate elements
$\si_{\pfrak}$ in $\rmGal(K[n]/{\mathbb Q})$ and to analyze its
characteristic polynomial. Consider a rational prime $p$ which is
different from $n$ and the conductor $N$, and therefore unramified
in $K[n]$, i.e. every prime ideal $\pfrak_i$ appears with
multiplicity one in the decomposition of $p$. Let $\pfrak$ be a
prime ideal in $K[n]$ which divides the principal ideal $(p)$.
Associated to these two ideals is the field extension
$\IF_{\rmN\pfrak}/\IF_p$, defined by the residue fields
$\IF_{\rmN\pfrak} = \cO_{K[n]}/\pfrak$ and $\IF_p = {\mathbb
Z}/(p)$. Here $\rmN\pfrak$ denotes the norm of the ideal $\pfrak$,
defined conceptually as the cardinality of the resulting residue
field. In the present context we are interested in Galois
extensions and the norm of a prime ideal can be computed most
easily by considering the Galois conjugates of the ideal \beq
\rmN\pfrak = \prod_{\si \in \rmGal(K[n]/{\mathbb Q})}
\si(\pfrak).\eeq  The Galois group of the finite extension
$\IF_{\rmN\pfrak}$ is particularly simple in that it is a cyclic
group whose generator is given by the Frobenius automorphism
$x\mapsto x^p$. Furthermore there is a surjective map from the
decomposition group, defined by \beq D_{\pfrak} = \{\si \in
\rmGal(K[n]/ {\mathbb Q})~|~\si(\pfrak) = \pfrak\}, \eeq onto this
Galois group \beq D_{\pfrak} \lra \rmGal(\IF_{\rmN\pfrak}/\IF_p).
\eeq The element $\si_{\pfrak}$ is defined as that element in the
decomposition group which corresponds to this generator of the
cyclic Galois group. Given the Frobenius automorphism of
$K[n]/{\mathbb Q}$ associated to $\pfrak$ one obtains the
following congruence \beq \rmdet(t\cdot {\bf 1} -
\rho(\si_{\pfrak})) = t^2 - c_pt + p ~{\rmmod}~n,\lleq{arithgal}
where $t$ is a formal variable and the coefficient $c_p$ is a
rational integer.

The connection with the arithmetic consideration is made by the
observation that the coefficients $c_p$ are determined by the
cardinalities of the curve. Results of this type are described in
\cite{s66}. It might appear that the Galois representations
contain only reduced information about the arithmetic of the
variety because of the mod condition in (\ref{arithgal}).
Surprisingly this is not the case if one considers not only
representations at any fixed prime $\ell$ but combines the
representations associated to $E[\ell^n]$, $n\in \IN$, into a
so-called $\ell$-adic representation $\rho_{\ell^{\infty}}$ of the
absolute Galois group $\rho_{\ell^{\infty}}: \rmGal({\bar {\mathbb
Q}}/{\mathbb Q}) \lra \rmGL(2,{\mathbb Z}_{\ell})$, where
${\mathbb Z}_{\ell}$ denotes the $\ell$-adic integers. Then one
has the following result.

{\bf Theorem.}\cite{jps68} {\it Each $\ell$-adic representation
determines the curve $E$ up to isogeny.}

The Galois theoretic framework suggests that an alternative way of
thinking about the reconstruction of spacetime from the string is
to ask what the Galois representations are that are induced by the
conformal field theory and how they are related to those that are
induced by the structure of spacetime. A number of results about
Galois representations that are either known or conjectured are of
relevance in this context.

It turns out that the mathematically easier direction is the one
that has been the more challenging from a physics perspective and
involves the problem of constructing spacetime from string theory.
Starting from a Hecke cusp eigenform $f$ of weight $k$ constructed
from a conformal field theory, results from Shimura, Deligne, and
Deligne-Serre show that there exists a continuous representation
\beq \rho_f:~\rmGal({\bar {\mathbb Q}}/{\mathbb Q}) \lra
\rmGL(2,\bIF_{\ell}) \eeq such that \bea \rmtr~\rho_f(\si_p) &=&
a_p~(\rmmod~\ell) \nn \\ \rmdet~\rho_f(\si_p) &=& \e(p)
p^{k-1}~(\rmmod~\ell),\eea where $p$ runs through the rational
primes not dividing $\ell$ and $\si_p\in \rmGal({\bar {\mathbb
Q}}/{\mathbb Q})$ is the Frobenius automorphism corresponding to
any prime ideal $\pfrak$ of $\bZZ$ lying over $p$. $\bIF_{\ell}$
is the algebraic closure of the finite field $\IF_{\ell}$ and $\e:
(\ZZ/N\ZZ)^* \lra \mathbb{C}^*$ is the character induced by the
diamond bracket operator (see e.g. \cite{rs00}). For modular forms
of higher weight $k$ the construction is involved, but for the
simpler case $k=2$, which is of relevance for the present context,
the Galois representation can be obtained via the Eichler-Shimura
theory described above. It emerges as the action of ${\rm
Gal}({\bar {\mathbb Q}}/{\mathbb Q})$ on the torsion points
obtained from an elliptic 'piece' of the modular curve $X_0(N)$.
More precisely, one combines all the torsion points into the Tate
module, an infinite tower of compatible torsion point groups on
which the Galois group acts.

Once we have a Galois representation, derived from the conformal
field theory, we can ask what the underlying geometry is of such
representations, if any. This line of thought is useful because a
result of Faltings \cite{gf83} shows that abelian varieties are
uniquely determined up to isogeny by their Galois representations.
This means that the arithmetic properties of abelian varieties are
determined by their Galois representation and explains why the
concept of modularity can be recovered in this context. Work in
this direction is based mostly on conjectures by Fontaine and
Mazur \cite{fm95}.

Fewer results are available in the direction which passes from the
geometry to modularity via the Galois group, because the
transition from Galois representations to modular forms is based
on conjectures by Serre which have yet to be proven.  Suppose we
start from a geometrically induced continuous irreducible
representation $ \rho:~\rmGal({\bar {\mathbb Q}}/{\mathbb Q}) \lra
\rmGL(2,\IF_{\ell})$ with odd determinant, i.e.
$\rmdet~\rho(\si_{\infty}) =-1$, where $\si_{\infty} \in
\rmGal({\bar {\mathbb Q}}/{\mathbb Q})$ is the restriction to
${\bar {\mathbb Q}}$ of complex conjugation in $\mathbb{C}$. Then
Serre's conjecture states that such a Galois representation is
induced by a modular form and it provides a prescription for
computing the modular level $N$, the weight $k$, and the character
$\e$ for which the eigenform $f$ should exist. A review of the
status of the Serre conjectures can be found in ref. \cite{rs00}.

\vskip .3truein

{\Large {\bf Acknowledgement}} \hfill \break It is a pleasure to
thank Monika Lynker and John Stroyls for discussions. Part of this
work was completed while R.S. was supported as a Scholar at the
Kavli Institute for Theoretical Physics in Santa Barbara. It is a
pleasure to thank the KITP and Indiana University at South Bend
for hospitality. This work was supported in part by the National
Science Foundation under Grant No. PHY99-07949.

\end{document}